\documentclass[showpacs,showkeys,prd,nofootinbib]{revtex4}
\usepackage{amsmath,graphicx}

\begin{document}

\title{Rho meson properties from combining QCD-based models}


\author{Stefan Leupold}


\affiliation{Institut f\"ur Theoretische Physik, Universit\"at
Giessen, D-35392 Giessen, Germany}

\begin{abstract}
Aiming at the calculation of the properties of $\rho$-mesons,
non-perturbative QCD-based methods are discussed concerning their potentials
as well as their short-comings. The latter are overcome by combining these techniques.
The utilized methods are (i) the chiral constituent quark model deduced from the 
instanton vacuum model and large-$N_c$ arguments, 
(ii) chiral perturbation theory unitarized by the
inverse amplitude method and (iii) QCD sum rules. Advantages of the combination of
these methods are especially the absence of un-physical quark-production thresholds and 
parameter-free results. Already in the chiral limit and
in leading order in $1/N_c$ one obtains a reasonable result for the mass of the 
$\rho$-meson, namely $m_\rho = 790 \pm 30 \,$MeV. Using the KSFR relation the
universality of the $\rho$-meson coupling is recovered. 
The latter is found to be $g = 6.0 \pm 0.3$.
\end{abstract}

\maketitle

\section{Introduction}
\label{sec:intro}

Two of the most interesting non-perturbative aspects of the strong interaction are 
chiral symmetry breaking ($\chi$SB) and confinement. 
It is neither fully settled how they come
about in QCD nor how important they are for the properties of a given hadronic species.
In the following we will consider the $\rho$-meson which is made of light quarks. 
The aim is to determine its mass as well as its coupling to pions and (virtual) photons. 
To describe the properties of mesons purely made out of {\em heavy} quarks it is quite
clear that confinement is essential while chiral symmetry is unimportant. 
One may speculate whether it is (more or less) the other way round for mesons 
purely made out of light quarks. This is supported by the following considerations:
Quark models show
that the mechanism of $\chi$SB causes large constituent quark masses of the order
of 300 -- 400 MeV. Hence even without confinement the creation of a quark-antiquark
pair is rather expensive. Therefore the role of confinement for the 
description of light hadrons is at least diminished by the appearance of $\chi$SB 
\cite{Diakonov:1996zi}. This suggests that 
the properties of light hadrons are {\em quantitatively} determined by the effect
of $\chi$SB. In such a scenario confinement enters only {\em qualitatively}
by excluding non-white states and quark-antiquark thresholds. It is well known
that such a picture works very well for pions 
\cite{Vogl:1991qt,Hatsuda:1994pi,Klevansky:1992qe}. 
One reason why one does
not need confinement to describe the properties of pions can be found in the fact that 
the mass of these quasi-Goldstone bosons is much below the (constituent!)
quark-antiquark threshold. This is of course different for other types of mesons.
At first glance it seems that this messes up the line of reasoning given above. 
The point however is 
that mesonic resonances --- most prominently the $\rho$-meson --- leave a trace also in 
the low-energy region much below their 
pole masses by mediating e.g.~pion-pion interactions \cite{Donoghue:1989ed}. 
Hence the key idea is that 
on the one hand side ($\chi$SB aspect) one can describe the low-energy region reliably 
by a (chiral!) quark model (without confinement) --- as this region is far away from the 
quark-antiquark production threshold. On the other hand side (confinement aspect)
mesonic resonances are supposed to mediate the interactions in this low-energy
regime. By combining these informations it should be possible to determine
masses and coupling constants of mesonic resonances in terms of quark model expressions.
It is the purpose of the
present work to apply that picture to $\rho$-mesons. 
Note that the presented idea resembles to some extent the way in which the masses of 
$Z$- and $W$-bosons are related to the low-energy(!) Fermi theory of weak interactions
\cite{pesschr}.

It is nowadays common wisdom that 
the spontaneous breaking of chiral symmetry leads to the appearance of a Goldstone
boson octet. 
This causes a 
large gap in the excitation spectrum of the observed hadrons. The lowest pseudoscalar 
octet appears to be light\footnote{The meson masses deviate from zero on account of 
the (small) current quark masses which explicitly break the chiral symmetry.}
while all other hadrons are heavy. Therefore at low energies QCD reduces to an 
effective theory where only the pseudoscalar mesons appear which interact with each 
other and with external sources. Spontaneous chiral symmetry breaking also demands that 
the meson self-interaction vanishes with vanishing energy. Therefore a systematic 
expansion in terms of the derivatives of the meson fields is possible. These 
considerations lead to the effective lagrangian of chiral 
perturbation theory ($\chi$PT) \cite{gasleut1,gasleut2}
presented in the following for three light flavors:
\begin{equation}
  \label{eq:chipt}
{\cal L}_{\chi \rm PT} = {\cal L}_1 + {\cal L}_2 + \mbox{higher order derivatives}
\end{equation}
with
\begin{equation}
  \label{eq:chipt2}
{\cal L}_1 = {1\over 4} \,F_\pi^2 \,
{\rm tr} (\nabla_\mu U^\dagger \nabla^\mu U) + \ldots  \,,
\end{equation}
\begin{equation}
{\cal L}_2 =  L_1 [{\rm tr}(\nabla_\mu U^\dagger \nabla^\mu U)]^2 
+ L_2 {\rm tr}(\nabla_\mu U^\dagger \nabla_\nu U) 
      {\rm tr}(\nabla^\mu U^\dagger \nabla^\nu U) 
+ L_3 {\rm tr}(\nabla_\mu U^\dagger \nabla^\mu U
      \nabla_\nu U^\dagger \nabla^\nu U) 
+ \ldots  \,,
\label{eq:chipt4}
\end{equation}
where we have only displayed the terms which are relevant for later use, i.e.~the ones
which remain present once all external fields and explicit chiral symmetry breaking terms
are put to zero. In $U$ the
pseudoscalar meson fields are encoded. 
$F_\pi$ denotes the pion decay constant (in the chiral
limit). We refer to \cite{gasleut2} for further details. 

At low energies $\chi$PT yields an excellent description of the hadron phenomenology
(see e.g.~the reviews \cite{Ecker:1995gg,Pich:1995bw,Scherer:2002tk}). 
However, the approach is limited to
the low-energy regime. The reason for this limitation are just the resonances which
appear in the meson-meson scattering amplitudes. A derivative expansion like the
one present in the $\chi$PT lagrangian (\ref{eq:chipt}) can only give 
polynomials in the kinematic variables while a resonance appears as a pole.
Therefore to extend the applicability of $\chi$PT to the region of mesonic resonances
non-perturbative methods (resummations) are needed --- which, of course, introduce some 
model dependence (see \cite{pelaez02} and references therein). 
The demand for exact unitarity is a key element of such non-perturbative
methods to keep the model dependence as small as possible. In the following we use
the inverse amplitude method (IAM) to unitarize $\chi$PT \cite{pelaez02}. In this way 
we extend its 
applicability to the resonance region. The word ``extension'' implies that one only
uses the information obtained from the $\chi$PT lagrangian to describe the meson-meson
scattering data up to (including) the resonance region. This information, however,
is already determined from the low-energy data. This fits exactly to our
philosophy that the low-energy regime can be used to learn something about the 
resonances. As shown in \cite{pelaez02} the application of the IAM to $\chi$PT
yields very good results for the meson-meson scattering data up to 1.2 GeV especially
also reproducing seven mesonic resonances (including the $\rho$-meson we are interested
in). The low-energy coefficients $L_i$ where used as fit parameters to describe the data.
It turned out that the values for the $L_i$ obtained in that fit agreed very well
with the corresponding values obtained from the scattering lengths, i.e.~from pure
low-energy data. 

Concerning our resonance of interest, the IAM connects the properties
of the $\rho$-meson to the low-energy coefficients of $\chi$PT. However, we are aiming
at a connection of the $\rho$-meson properties to the underlying quark (and gluon) 
structure. As a purely hadronic model the IAM has the same limitation as pure
$\chi$PT: While the general structure of the $\chi$PT lagrangian is dictated by QCD,
the low-energy coefficients are pure fit parameters. Their connection to QCD is
not settled yet. To proceed we have to involve models with quark degrees of freedom.
This, of course, brings in a higher degree of model dependence. To some extent we
will relax that problem by using two independent approaches to cross-check our results.
In addition, we will choose methods which are not pure guesses but can be obtained
within well-defined approximations from QCD. To become specific we will use the chiral
constituent quark model proposed e.g.~in \cite{diakpet,espraf} and the QCD sum rule
approach \cite{shif79}. In the following we shall briefly comment on their merits but 
especially also on their short-comings (as we have done for unitarized $\chi$PT). 
Actually the latter are very specific for the respective
method. Therefore a key idea of the present work is to overcome the short-comings by 
combining all the approaches, i.e.~the IAM, the quark model and the sum rule approach.

The QCD sum rule method makes contact between the hadronic and the quark-gluon 
world \cite{shif79}. Two expressions for a correlation function are matched in the 
regime of deeply space-like momenta: the operator product expansion and a dispersion
integral utilizing the (hadronic) spectral representation of the correlation function.
The coefficients of the operator product expansion are given in terms of QCD perturbation
theory and quark and gluonic condensates, i.e.~by quark and gluon degrees of freedom. 
On the other hand, the spectral representation is given (at least in principle) 
in terms of hadronic degrees of freedom. Concerning the quark-gluon side, there are
condensates which are not or not fully known. They constitute a source of uncertainties 
for the sum rule method. One much debated issue is the question whether the four-quark
condensate is (approximately) given by the product of two-quark condensates 
(see e.g.~\cite{Leupold:1998dg} and references therein). Concerning the hadronic
side, there are ambiguities and model dependences in the spectral representation.
In principle, several hadronic states contribute to a given correlation function.
Typically one wants to learn something about the lowest lying state. This can only
be achieved if the higher lying ones are replaced by a simple expression with a minimal
number of free parameters \cite{shif79}. We will come back to that problem below.
But even for the lowest lying state several parameters are necessary to characterize it,
like peak position, peak height and peak width. As pointed out in 
\cite{klingl2,Leupold:1998dg} one cannot pin down all of these hadronic parameters from
the QCD sum rule method alone. 
As we will see below the combination of the sum rule technique with other methods will
allow for simplifications which diminish some of the mentioned uncertainties.

Quark models provide a microscopic approach to calculate hadronic properties in terms
of quark degrees of freedom. One rather generic feature of quark models, which we also use
in the following, is the expansion in terms of $1/N_c$ (see e.g.~\cite{Klevansky:1992qe})
where $N_c$ is the number of colors. A severe short-coming of quark models which
do not include confining forces is the appearance of unphysical decay channels
(concerning the $\rho$-meson cf.~e.g.~\cite{He:1998gn}). On the other hand, confining
quark models usually have non-local lagrangians or contain non-relativistic parts.
Since we want to study in the present work how far we can get without a quantitative 
influence of confinement (cf.~discussion above) we will not comment on confining
quark models any more. A further problem of most quark models is the appearance 
of a couple of free parameters which have to be fitted to some hadronic observables.
This clearly limits the predictive power. Finally we note as a further short-coming
that the connection of some quark models to QCD is not clear. 

In the present work we will utilize a chiral constituent quark model which can be 
deduced \cite{diakpet} from the instanton model of QCD \cite{schaefshur}
using large-$N_c$ arguments. It is
actually the simplest chiral model one can write down which couples quarks and Goldstone 
bosons. Therefore it has a minimal number of free parameters. Even more, for the
quantities we will need in the following the results are parameter-free.

All the methods discussed above which we will use in the following constitute
well-established techniques. The basic new idea of the present work is the combination
of these methods. The purpose is to avoid some of the short-comings which appear
if only one the methods is used alone. Especially the use of the quark model is
restricted to the low-energy region where the unphysical quark-antiquark decay channel
is far away. Low-energy coefficients of $\chi$PT are calculated from the quark
model. The applicability of the obtained $\chi$PT lagrangian is extended to higher
energies using the IAM. At this stage the model is purely hadronic, i.e.~no unphysical
thresholds appear. The IAM generates the $\rho$-meson, i.e.~connects the $\rho$-meson
properties to the low-energy coefficients. The latter in turn have been connected to 
quark degrees of freedom by the quark model. Finally the QCD sum rule method is
used to further justify the obtained results and to get an error estimate especially
for the quark model calculations. We want to note here that we will not try to
estimate the error induced by the use of the IAM. This would require the knowledge
of higher order terms in the $\chi$PT lagrangian which are not fully available at 
present.

As already pointed out the quark model we will use works in leading order in a
$1/N_c$-expansion. Therefore to be consistent we will have to analyze and use also
the other methods in that large-$N_c$ limit. 
As we will see below this apparent restriction will actually induce 
very welcome simplifications for the IAM and the sum rule approach.
We will also work in the chiral limit for simplicity. For the $\rho$-meson
made out of light up- and down-quarks we do not
expect drastic deviations by that simplification. In addition, it is actually in this
limit where the quark model calculations yield parameter-free results. We will
come back to that point below.

The paper is organized in the following way. In the next section we review briefly the
calculation of the chiral low-energy coefficients from the chiral constituent quark
model. Sec.~\ref{sec:IAM} contains a pedagogical example which introduces the IAM.
We also present there the application of the IAM to $\chi$PT in the combined large-$N_c$
and chiral limit. Combined with the quark model results from Sec.~\ref{sec:quarkmod}
the $\rho$-meson properties are calculated. Sec.~\ref{sec:QCDSR} presents the 
QCD sum rule method. Together with the IAM results an error estimate for the
previous calculations are obtained. In Sec.~\ref{sec:res} the results are compared
to experiments and to other approaches in the literature. Finally we summarize
in Sec.~\ref{sec:sum} and give an outlook how the presented method can be extended.

\section{Chiral constituent quark model}
\label{sec:quarkmod}

The chiral constituent quark model is defined by the 
lagrangian \cite{diakpet,espraf}
\begin{equation}
  \label{eq:lagrconst}
{\cal L}_{\rm quark} = 
\bar q \, (\gamma_\mu\partial_\mu - M U^{\gamma_5} + \ldots)\,q
\end{equation}
with the matrix for the Goldstone meson fields
\begin{equation}
  \label{eq:defUg5}
  U^{\gamma_5} = {1-\gamma_5 \over 2} \, U + {1+\gamma_5 \over 2} \, U^\dagger
\end{equation}
and the constituent quark mass $M$.
The dots in (\ref{eq:lagrconst}) denote couplings to external sources and explicit 
chiral symmetry breaking terms which we do not
need here.\footnote{The latter we do not need as we work in the chiral limit.}
As shown in \cite{diakpet}
the lagrangian (\ref{eq:lagrconst}) can be derived as the low energy limit of the 
instanton model. It is correct up to corrections which are suppressed by the number
of colors $N_c$ and/or by the ratio of size and mean distance of instantons.
In \cite{espraf} the same lagrangian is derived on different grounds from the QCD
lagrangian by symmetry considerations concerning the axial anomaly, large-$N_c$
arguments and neglecting gluonic contributions.
In any case, (\ref{eq:lagrconst}) is the simplest model with Goldstone boson and
constituent quark degrees of freedom. We note in passing that this lagrangian
is also used for the description of nucleons as non-topological solitons 
(cf.~\cite{Christov:1996vm} and references therein). 

\begin{figure}[hbt]
\includegraphics[keepaspectratio,width=5cm]{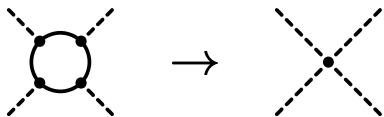}
  \caption{Schematic view of the low-energy reduction of the chiral constituent
quark model to $\chi$PT. The dashed
lines denote Goldstone bosons and the full lines constituent quarks. 
Note that the diagram on the 
l.h.s.~stands for the sum of all possible diagrams of that type (box diagrams). This
includes also diagrams with crossed quark lines.}
  \label{fig:fourpoint}
\end{figure}

The lagrangian in 
(\ref{eq:lagrconst}) is given in Euclidean space. As instantons are classical solutions
of the Euclidean Yang-Mills equations \cite{schaefshur} the lagrangian 
(\ref{eq:lagrconst}) is derived from the Euclidean QCD action using the instanton 
model \cite{diakpet}. We hesitate to transform (\ref{eq:lagrconst}) to Minkowski space
as it would introduce the already mentioned unphysical quark-antiquark production
thresholds due to the lack of confinement. We will come back to that point below.
By integrating out the quarks, expanding the obtained effective action in terms of meson
field derivatives and finally transforming the result to Minkowski space
one arrives at the $\chi$PT lagrangian (\ref{eq:chipt}) with predictions for the 
low-energy constants. This procedure is shown schematically in Fig.~\ref{fig:fourpoint}.
For the low-energy constants 
one obtains \cite{diakpet,espraf}
\begin{subequations}
\label{eq:quark}
\begin{eqnarray}
\label{eq:quark1}
L_1 & = & {1 \over 24} \, {N_c \over 16 \pi^2}  \,,
\\  
\label{eq:quark2}
L_2 & = & {1 \over 12} \, {N_c \over 16 \pi^2} \,,
\\
\label{eq:quark3}
L_3 & = & -{1 \over 6} \, {N_c \over 16 \pi^2} \,.
\end{eqnarray}
\end{subequations}
We note again that these expressions are correct in leading order of a $1/N_c$ 
expansion. We also note that the coefficients $L_i$ are in general scale dependent 
quantities \cite{gasleut1,gasleut2}. 
This dependence, however, is subleading in the $1/N_c$ expansion
and therefore of no relevance for our purposes. In addition, as we will see
in the next section, we shall be merely interested in $L_3$. This specific quantity
is scale independent. 
It is an important aspect of the used quark model that the results (\ref{eq:quark}) 
are {\em parameter-free}. This, of course, it not true for all low-energy constants.
In principle, the lagrangian (\ref{eq:lagrconst}) explicitly contains the constituent
quark mass as an input parameter and in addition a UV-cutoff prescription as the model
is not renormalizable. If the quark model is derived from the instanton vacuum model
both mass and UV-cutoff can be related to the instanton properties \cite{diakpet}.
The reason why $L_1$ -- $L_3$ are pure numbers lies in the fact that the sum of the
box diagrams depicted in Fig.~\ref{fig:fourpoint} is UV-finite, i.e.~insensitive to 
the UV-cutoff.
On the other hand, being dimensionless quantities the $L_i$'s cannot depend solely
on the constituent quark mass which presents the only other dimensionful scale in the 
model. This is the deeper reason for the parameter-free results obtained in 
(\ref{eq:quark}). Of course, also the constant $F_\pi$ which appears in 
(\ref{eq:chipt2}) can be obtained by integrating out the quarks from 
(\ref{eq:lagrconst}). This low-energy constant, however, is sensitive to the UV-cutoff
\cite{diakpet,espraf}. Instead of specifying the UV-cutoff prescription such that
the desired physical result for $F_\pi$ is obtained we simply will use the physical
value for $F_\pi$ as an input.

Since the presented quark model is restricted to the leading order in $1/N_c$ we have
to enforce this restriction also upon the other methods we are using subsequently.
At first sight one might regard that as a short-coming.
On the other hand, this restriction has also a positive aspect: As we shall see below 
it keeps the utilized methods rather simple. Anyway we can study how far we can get
within that restriction. For arguments involving $N_c$-counting it is important to
note that $F_\pi^2$ is of order $N_c$ \cite{espraf}.

\section{Inverse amplitude method}
\label{sec:IAM}

The application of the inverse amplitude method (IAM) to $\chi$PT is discussed in 
detail in \cite{Dobado:1997ps,pelaez02} (see also earlier references cited therein). 
Nonetheless, to keep our work selfcontained 
we will present here a very general introduction to the IAM. The purpose is to 
show how a resonance leaves its trace in the low-energy region much below its pole
mass (cf.~the discussion in the introduction and e.g.~\cite{Donoghue:1989ed}) 
and how in turn
the low-energy scattering amplitude can be used to ``(re)generate'' the resonance.
A pedagogical example \cite{Oller:1998hw}
will serve to introduce the general idea of the IAM before
we come back to our specific application of this technique. A central aspect
of the whole method is unitarity, i.e.~the condition $S \, S^\dagger = 1$ for the
$S$-matrix. In the following we only need one channel. In this case, $S$ is related
to the scattering amplitude via \cite{pelaez02}
\begin{equation}
  \label{eq:sandt}
S = 1 + 2 i \sigma T  \,.
\end{equation}
Here $\sigma = 2 k/\sqrt{s}$ where $k$ is the momentum of the involved particles in the
center-of-mass system. From the unitarity requirement one can deduce
\begin{equation}
  \label{eq:imtm1}
{\rm Im} T = \sigma \vert T \vert^2 
\qquad \Rightarrow \qquad 
{\rm Im} T^{-1} = -\sigma   \,,
\end{equation}
i.e.~the imaginary part of the {\em inverse} scattering amplitude is completely
fixed by unitarity. 

Next we introduce the following pedagogical example \cite{Oller:1998hw}:
Suppose the scattering of two particles (for simplicity with the same mass $m$)
is mediated by a resonance (with mass $M$) in the p-wave channel. Hence the scattering 
amplitude in our toy model is given by 
(we neglect the real part of the resonance self energy)
\begin{equation}
  \label{eq:resscattoy}
T = - {g^2 k^2 \over 12 \pi} {1 \over s-M^2 + i {\rm Im}\Pi(s) }
\end{equation}
with the center-of-mass momentum of the scattering particles
\begin{equation}
  \label{eq:defktoy}
k = \sqrt{ {s \over 4} - m^2}
\end{equation}
and a coupling constant $g$. For the (imaginary part of the) self energy of the
resonance one gets (we restrict ourselves to a single channel problem)
\begin{equation}
  \label{eq:impitoy}
{\rm Im}\Pi(s) = {g^2 \over 6 \pi} {k^3 \over \sqrt{s} }  \,.
\end{equation}
We note that this scattering amplitude fulfills the unitarity relation (\ref{eq:imtm1})
with the available phase space $\sigma = 2 k/\sqrt{s}$.

For energies (much) below the resonance mass $M$ one can expand the scattering amplitude
as a Taylor series in powers of $1/M^2$. The lowest order contribution is given by
\begin{equation}
  \label{eq:t2toy}
T_2 = {g^2 k^2 \over 12 \pi M^2}  \,.
\end{equation}
This is a valid approximation for the scattering amplitude $T$ for {\em very} low 
energies. For somewhat higher energies (but still in the low-energy region much
below the resonance mass) one can improve this approximation by calculating the
$o(1/M^4)$-contribution:
\begin{equation}
  \label{eq:t4toy}
T_4 = {g^2 k^2 \over 12 \pi M^4} (s + i {\rm Im}\Pi(s) )
\end{equation}
especially
\begin{equation}
  \label{eq:ret4toy}
{\rm Re}T_4 = T_2 \, {s \over M^2}
\end{equation}
where the last expression is suitable for later use.
It is important to note that the Taylor series has a finite radius of convergence.
E.g.~for a sharp resonance (Im$\Pi \approx 0$) the series converges for $s \le M^2$.
Therefore the low-energy approximation
\begin{equation}
  \label{eq:tt2t4}
T \approx T_2 + T_4
\end{equation}
cannot give a reliable estimate for higher energies as
the Taylor expansion breaks down for energies around the resonance mass.
The reason for this breakdown simply is the appearance of the resonance pole in
the scattering amplitude (\ref{eq:resscattoy}) which cannot be described by a
polynomial in $1/M^2$. The basic assumption of the IAM is that 
the {\em inverse amplitude} has a larger radius of convergence as compared 
to the amplitude itself. We recall that due to unitarity the imaginary part of the
inverse amplitude is always fixed by (\ref{eq:imtm1}). Hence what remains to be
done is to calculate the real part of the inverse amplitude. Again we perform that
calculation at low energies, but expect that the range of application for this
approximate inverse scattering amplitude is larger than the range of application
for the original low-energy scattering amplitude. For our toy model we get
\begin{equation}
  \label{eq:invtoy}
T^{-1} = (T_2 + T_4 + o(1/M^6) \,)^{-1} = 
T_2^{-1} (T_2 - T_4  + o(1/M^6) \,) T_2^{-1}  \,.
\end{equation}
As $T_2$ is purely real we get for the real part
\begin{equation}
  \label{eq:invrealtoy}
{\rm Re}T^{-1} \approx  T_2^{-1} (T_2 - {\rm Re}T_4 ) T_2^{-1}  \,.
\end{equation}
Finally we obtain the following relation which is actually not restricted to our
toy model but generally valid \cite{pelaez02}:
\begin{equation}
  \label{eq:tinvbasform}
T \approx T_2 \, (T_2 - {\rm Re}T_4 - i T_2 \sigma T_2)^{-1}\, T_2   \,.
\end{equation}
For our case at hand we get
\begin{equation}
  \label{eq:iamrestoy}
T(s) = T_2 \left( 1 - {s \over M^2} - i T_2 \sigma \right)^{-1}
= -{ T_2 M^2 \over s - M^2 + i T_2 M^2 \sigma }
\end{equation}
which completely agrees with the resonant amplitude (\ref{eq:resscattoy}).
In general, (\ref{eq:tinvbasform}) is (only) an approximation for the true scattering
amplitude as higher order terms $T_6, \ldots$ were neglected. 
The simplicity of the chosen toy model (\ref{eq:resscattoy}) allowed
for a complete reconstruction via the first two low energy amplitudes. In general,
realistic scattering amplitudes are not as simple as the presented toy model. In
particular, crossing symmetry typically yields additional contributions besides the 
$s$-channel resonance of (\ref{eq:resscattoy}). For a discussion of these issues we
refer to \cite{Dobado:1997ps,Boglione:1997uz,pelaez02}.

It is important to note that the input for the result (\ref{eq:tinvbasform}) solely
comes from the low-energy amplitudes (\ref{eq:t2toy}) and (\ref{eq:ret4toy})
which are only valid below the resonance region. In practice the situation is
such that one knows the low-energy amplitudes but does not know the amplitude in
the resonance region. Using (\ref{eq:tinvbasform}) one obtains an expression
which is supposed to be valid also in the resonance region.
We conclude that the IAM is capable of determining resonance properties from the
low-energy appearance of the resonance. Hence this technique fits exactly to
our philosophy outlined in the introduction: The quark model should only be trusted in
the low-energy region far away from the unphysical quark-antiquark thresholds. This
region, however, can also be used to learn something about hadronic resonances.
The IAM provides a tool to translate this idea into a quantitative calculation.

Now we turn to the resonance we are actually interested in. The $\rho$-meson
appears as a resonant state in pion-pion scattering. Indeed one can construct the
$\rho$ from the low-energy pion scattering amplitude in the vector--iso-vector channel 
using the IAM (cf.~\cite{Dobado:1997ps,pelaez02} and references therein).
Strictly speaking concerning the scattering of members of the Goldstone boson octet
the vector--iso-vector channel constitutes a two-state problem, namely $\pi\,\pi$
and $K \bar K$ appear in that channel. For realistic masses the $K \bar K$-production
threshold is above the $\rho$-meson mass. Therefore, it should be possible to
discuss the $\rho$-properties by studying only elastic pion-pion scattering. Indeed, this
is what we will do in the following. On the other hand, however, since we work in the
chiral limit for the whole octet, the decay of a $\rho$-meson into massless kaons
is of course possible in our approach. We have checked that the treatment of the
two-state problem within the IAM yields the same result as the one-state problem
which we will present in the following.\footnote{Of course, the {\em total} width of the
$\rho$ is larger for the two-state approach as it accounts for decays both into pions 
and kaons. The partial decay width into pions agrees with the one presented here.}

Within the IAM the scattering amplitude is given by (\ref{eq:tinvbasform}).
In the combined chiral and large-$N_c$ limit the relevant expressions \cite{pelaez02} 
for the calculation of the pion scattering amplitude in the vector--iso-vector channel 
become rather simple:
\begin{equation}
  \label{eq:sigmaeq1}
\sigma = 1   \,,
\end{equation}
\begin{equation}
  \label{eq:t2}
T_2(s) = {s \over 96 \pi F_\pi^2}  \,,
\end{equation}
\begin{equation}
  \label{eq:ret4}
{\rm Re}T_4(s) = - {L_3 s^2 \over 24\pi F_\pi^4} + o(1/N_c^2) 
= -{4 L_3 s \over F_\pi^2} \, T_2(s) + o(1/N_c^2) 
\,.
\end{equation}
Thus we get for the (resonant) scattering amplitude
\begin{equation}
  \label{eq:finscattamp}
T(s) = T_2(s) \, \left[ 1+ {4 L_3 s \over F_\pi^2} - i T_2(s) \right]^{-1}
= {-{\rm Im}\Pi(s) \over s - M_V^2 + i {\rm Im}\Pi(s) }
\end{equation}
with the resonance mass (squared)
\begin{equation}
  \label{eq:mass1}
M_V^2 := - {F_\pi^2 \over 4 L_3}
\end{equation}
and the imaginary part of the resonance self energy 
\begin{equation}
  \label{eq:width}
{\rm Im}\Pi(s) := -{s \over 24 \cdot 16 \pi L_3}
\end{equation}
Note that ${\rm Im}\Pi$ is of order $1/N_c$, i.e.~vanishes in the large-$N_c$ limit
as it should be, as mesonic (quark-antiquark) resonances become stable
in that limit \cite{witten}. Hence also the real part of the self energy 
is of order $1/N_c$, i.e.~subleading as compared to $M_V^2$. As we have only
presented the leading order for ${\rm Re}T_4$ in (\ref{eq:ret4}) we have consequently 
not displayed the real part of the self energy in (\ref{eq:finscattamp}).
The imaginary part ${\rm Im}\Pi$ contains the information about the coupling of the
generated resonance to the pions. To extract a coupling constant (i.e.~a number) we
need a model lagrangian for $\rho$-mesons and pions. We take the
standard $\rho\pi\pi$-lagrangian \cite{klingl1}
\begin{equation}
  \label{eq:standlagr}
{\cal L}_{\rm int} = {ig \over 4}\, {\rm tr}(V^\mu \, [\partial_\mu \Phi,\Phi])
- {g^2 \over 16}\, {\rm tr}([V^\mu,\Phi]^2)
\end{equation}
where $V^\mu$ denotes the $\rho$-meson and $\Phi$ the pion field. Calculating
${\rm Im}\Pi$ from (\ref{eq:standlagr}) in the chiral limit in the one-loop 
approximation 
one 
obtains\footnote{Note that our convention for the self-energy differs from the one
in \cite{klingl1} by a sign, cf.~also (\ref{eq:fullrhoprop}) below.}
\begin{equation}
  \label{eq:impistdmod}
{\rm Im}\Pi(s) = {g^2 s \over 48 \pi}  \,.
\end{equation}
Comparing (\ref{eq:width}) with (\ref{eq:impistdmod}) yields
\begin{equation}
  \label{eq:gdet}
g^2 = - {1 \over 8 L_3}  \,.
\end{equation}

Finally we can determine the coupling $g_\gamma$ of the $\rho$-meson to an 
electromagnetic current, i.e.~to (virtual) photons. Chiral symmetry demands the KSFR
relation \cite{ks,fr} to hold:
\begin{equation}
  \label{eq:ksfr}
M_V^2 = 2 g g_\gamma F_\pi^2  \,.
\end{equation}
Using (\ref{eq:mass1}) and (\ref{eq:gdet}) we get
\begin{equation}
  \label{eq:ggammadet}
g_\gamma = {M_V^2 \over 2 g F_\pi^2} = {M_V^2 \over 2 g^2 F_\pi^2} \, g = g  \,,
\end{equation}
i.e.~we have obtained the universality of the $\rho$-meson coupling. We note in passing
that this strict universality does not hold any longer once we leave the chiral limit. 
This, however, is beyond the scope of the present work. The fact that
the $\rho$-meson couples (at least approximately) with the same strength to
pions and photons constitutes one important aspect of the vector meson dominance
picture (e.g.~\cite{klingl1,herrmann,Dung:1996rp}, 
see also comment in \cite{Ecker:1989yg}). 

Note that so far we have not used any input from the quark model calculations of
Sec.~\ref{sec:quarkmod} {\em except} that we considered only the leading order of 
the large-$N_c$ expansion. Hence, the obtained universality of the $\rho$-meson
coupling is {\em not} a result of a specific quark model but of the IAM treated in the
large-$N_c$ approximation and in the chiral limit.

In view of the results (\ref{eq:mass1}), (\ref{eq:gdet}) and (\ref{eq:ggammadet})
the only missing piece is an expression for $L_3$. 
Clearly this cannot be obtained from the IAM as a purely hadronic approach.
Next we use the results of Sec.~\ref{sec:quarkmod} to determine $L_3$ and fix
in that way
mass and coupling constant of the $\rho$-meson. Alternatively we will use in the
next section the QCD sum rule approach to determine $L_3$. We will see that we obtain
a rather consistent picture within the two approaches (quark model and QCD sum rules).
Using (\ref{eq:quark3}), (\ref{eq:mass1}), (\ref{eq:gdet}) and (\ref{eq:ggammadet})
yields the following results obtained from a combination of the quark model with
the IAM:
\begin{subequations}
\label{eq:qmIAM}
\begin{eqnarray}
  \label{eq:qmIAMmass}
M_V & = & \sqrt{{3 \over N_c}} \, \sqrt{8} \pi F_\pi \,,
\\
\label{eq:qmIAMcoupl}
g = g_\gamma & = & \sqrt{{3 \over N_c}} \, 2 \pi  \,.
\end{eqnarray}
\end{subequations}
We will confront these results with experimental findings in Sec.~\ref{sec:res}
after an alternative determination of $L_3$, $M_V$ and $g$ using QCD sum rules in
the next section.

\section{QCD sum rules}
\label{sec:QCDSR}

The starting point of this method is the
covariant 
time ordered current-current correlator 
\begin{equation}
\Pi_{\mu\nu}(q) = i \int\!\! d^4\!x \, e^{iqx} 
\langle T j_\mu(x) j_\nu(0) \rangle    \,.
  \label{eq:curcur}
\end{equation}
For the $\rho$-meson channel, $j_\mu$ is the isospin-1 part of the electromagnetic 
current,
\begin{equation}
\label{eq:curud}
j_\mu = {1\over 2} \left( \bar u \gamma_\mu u - \bar d \gamma_\mu d \right)   \,.
\end{equation}
Due to Lorentz invariance and charge conservation the correlator (\ref{eq:curcur}) 
has the following decomposition \cite{shif79}
\begin{equation}
  \label{eq:decompmunu}
\Pi_{\mu\nu}(q) = \left( q_\mu q_\nu - g_{\mu\nu} q^2 \right) R(q^2)  \,.
\end{equation}
In the time-like region, $q^2 > 0$, the quantity $R$ is related to the cross section 
$e^+ e^- \to $ hadrons with isospin 1 via
\cite{pastar}
\begin{equation}
  \label{eq:crosssec}
{ \sigma^{I=1}(e^+ e^- \to \mbox{hadrons}) \over \sigma(e^+ e^- \to \mu^+ \mu^-) }
= 12 \pi {\rm Im}R   \,.
\end{equation}
At least for low energies the time-like region is determined by hadronic degrees 
of freedom. 
We denote the result for $R$ in the 
time-like region by $R^{{\rm HAD}}$. On the other hand, the current-current 
correlator (\ref{eq:curcur}) can be calculated for $q^2 \ll 0$ 
using Wilson's operator product expansion (OPE) \cite{wilson69} 
for quark and gluonic degrees of freedom \cite{shif79}.
In the following we shall call the result 
of that calculation $R^{{\rm OPE}}$.
A second representation in the space-like region which has to match 
$R^{{\rm OPE}}$ can be obtained from $R^{{\rm HAD}}$ by
utilizing a subtracted dispersion relation. We find 
\begin{eqnarray}
R^{{\rm OPE}}(Q^2) & = & \tilde c 
-{Q^2 \over \pi} \int\limits_0^\infty \!\! ds \,
{{\rm Im}R^{{\rm HAD}}(s) \over (s+Q^2)s}
  \label{eq:opehadr}
\end{eqnarray}
with $Q^2:= -q^2 \gg 0$ and a subtraction constant $\tilde c$.

For the l.h.s.~of (\ref{eq:opehadr}) the OPE yields in the combined chiral and
large-$N_c$ limit\footnote{Note that only in the large-$N_c$ limit there is a
strict connection between the four-quark condensate appearing in the OPE and 
the two-quark condensate \cite{Novikov:1984jt}.} \cite{shif79}
\begin{equation}
  \label{eq:decconR}
R^{{\rm OPE}}(Q^2) = 
-c_0 \,{\rm ln}\left({Q^2 \over \mu^2}\right) + {c_1 \over Q^2}
+ {c_2 \over Q^4} + {c_3 \over Q^6} + o(1/Q^8,\alpha_s^2)
\end{equation}
with
\begin{subequations}
\label{eq:coeffrho}
\begin{eqnarray}
\label{eq:c0def}
c_0 & = & {N_c \over 24 \pi^2} \, \left( 1 + {\bar\alpha_s \over \pi} \right)  \,,  \\
c_1 & = & 0  \,,  \\
c_2 & = & {1\over 24} \left\langle {\alpha_s \over \pi} G^2 \right\rangle     \,,  \\
\label{eq:c3def}
c_3 & = & -{112 \over 81} \pi\bar\alpha_s {3 \over N_c} \, 
\langle \bar q q \rangle^2  \,.
\end{eqnarray}
\end{subequations}
The actual values we use for the strong coupling constant $\alpha_s$, the
gluon condensate $\left\langle {\alpha_s \over \pi} G^2 \right\rangle $ and
the quark condensate $\langle \bar q q \rangle $ will be discussed in Sec.~\ref{sec:res}.
We have introduced a quantity $\bar\alpha_s$ which becomes identical to $\alpha_s$ for
three colors but remains finite for $N_c \to \infty$:
\begin{equation}
  \label{eq:alphasbar}
  \bar\alpha_s = {N_c^2 -1 \over N_c} \, {3 \over 8} \, \alpha_s  \,.
\end{equation}
Note that in the spirit of the $1/N_c$-expansion the strong coupling constant
$\alpha_s$ is $o(1/N_c)$. 

Concerning the r.h.s.~of (\ref{eq:opehadr}) we note that in the large-$N_c$ approximation
${\rm Im}R^{{\rm HAD}}$ is given by the spectral representation of
(infinitely many) {\em stable} mesons \cite{witten}:
\begin{equation}
  \label{eq:ressum}
{\rm Im}R^{{\rm HAD}}(s) = \pi \, {M_V^2 \over g_\gamma^2} \, \delta(s-M_V^2)
+ \sum\limits_{i = 2}^\infty   \pi \, {M_i^2 \over g_i^2} \, \delta(s-M_i^2)
\end{equation}
where we have isolated the lowest-lying meson we are actually interested in. $g_i$
denotes the coupling constant between the photon field and the respective meson.
Clearly with such a representation with infinitely many mesons, i.e.~infinitely
many {\it a priori} unknown parameters, we cannot learn 
anything about the $\rho$-meson properties. To simplify the expression we invoke
a duality argument: For high energies
the quark structure of the current-current correlator is resolved. QCD perturbation
theory becomes applicable yielding \cite{shif79}
\begin{equation}
  \label{eq:qcdperthe}
{\rm Im}R^{{\rm HAD}}(s) 
=  \pi c_0   \qquad 
\mbox{for large $s$,}
\end{equation}
where the constant $c_0$ is given in (\ref{eq:c0def}).
These considerations suggest to approximate the sum of higher-lying resonances
by a continuum starting at some threshold $s_0$ which separates the low- from the
high-energy region:
\begin{eqnarray}
{\rm Im}R^{{\rm HAD}}(s) 
& = &
\Theta(s_0 -s) \, {\rm Im}R^{{\rm RES}}(s) 
+ \Theta(s -s_0) \, \pi c_0    \,.
  \label{eq:pihadans}
\end{eqnarray}
The mesons below the threshold are now collected in ${\rm Im}R^{{\rm RES}}$. Later
we will use only the $\rho$-meson, i.e.~the lowest-lying one, here. For the
moment we want to keep the expression more general.
Of course, the high-energy behavior given in (\ref{eq:pihadans}) is only an 
approximation on the true spectral distribution for the current-current 
correlator. Duality states that taking appropriate energy integrals yields
the same results no matter whether one takes (\ref{eq:ressum}) or (\ref{eq:pihadans})
to represent the current-current correlator. The phrase ``appropriate energy integrals''
will be further qualified below. We note that the duality argument
becomes even more convincing when changing from large-$N_c$ considerations to the
real world of three colors. There the mesons really become resonances with sizable
widths. The higher-lying resonances more or less dissolve into a continuum which indeed
has the height as predicted by perturbative QCD (cf.~e.g.~Fig.~1 in \cite{klingl2}).

Next we insert the simplified representation (\ref{eq:pihadans}) in the dispersion
relation (\ref{eq:opehadr}) and expand also the r.h.s.~in powers of $1/Q^2$. In this
way one can compare
the coefficients of this expansion with the respective ones in the series on the 
l.h.s.~given by (\ref{eq:decconR}). This yields the finite energy sum 
rules\footnote{This derivation is actually oversimplified since it neglects the running
of the coupling constant. For a rigorous derivation cf.~\cite{maltman} and references
therein.}
\begin{subequations}
\label{eq:fesr}
\begin{eqnarray}
\label{eq:fesr1}
{1 \over \pi} \int\limits_0^{s_0} \!\! ds \, {\rm Im}R^{{\rm RES}}(s)
- c_0 s_0 
& = & 0   \,,
\\  
\label{eq:fesr2}
-{1 \over \pi} \int\limits_0^{s_0} \!\! ds \, s \,{\rm Im}R^{{\rm RES}}(s)
+ c_0 \, {s_0^2 \over 2} 
& = & c_2   \,,
\\
\label{eq:fesr3}
{1 \over \pi} \int\limits_0^{s_0} \!\! ds \, s^2 \,{\rm Im}R^{{\rm RES}}(s)
- c_0 \, {s_0^3 \over 3} 
& = & c_3  \,,
\end{eqnarray}
\end{subequations}
where the coefficients of the OPE are given in (\ref{eq:coeffrho}). Note that the
powers of $s$ in the respective integrand rise from (\ref{eq:fesr1}) to
(\ref{eq:fesr3}). Hence the higher-order sum rules become more sensitive to details of 
the parametrization in the high-energy region. In view of our simplified representation
(\ref{eq:pihadans}) as compared to the exact one (\ref{eq:ressum}), one has to
be aware of the fact that the finite energy sum rules become less reliable from
(\ref{eq:fesr1}) to (\ref{eq:fesr3}). We will come back to that point below.
Obviously the sum rules which (approximately) hold constitute a set of 
``appropriate energy integrals'' in the sense of the duality picture discussed above.

In the large-$N_c$ approximation the parametrization of the spectral information
for the $\rho$-meson is given by
\begin{equation}
  \label{eq:pararho}
{\rm Im}R^{\rm RES}(s) = \pi \, {M_V^2 \over g_\gamma^2} \, \delta(s-M_V^2)  \,.
\end{equation}
From the results (\ref{eq:mass1}), (\ref{eq:gdet}) and (\ref{eq:ggammadet})
of Sec.~\ref{sec:IAM} we deduce
\begin{equation}
  \label{eq:pararho2}
{\rm Im}R^{\rm RES}(s) = 2 \pi F_\pi^2 \, \delta(s-M_V^2)  \,.
\end{equation}
We can use the first sum rule (\ref{eq:fesr1}) to fix $s_0$:
\begin{equation}
  \label{eq:dets0sr}
s_0 = {2 F_\pi^2 \over c_0}
= {3 \over N_c} \, {(4 \pi F_\pi)^2 \over 1 + {\bar\alpha_s \over \pi} }
\end{equation}
and the second one (\ref{eq:fesr2}) to determine $M_V^2$:
\begin{equation}
  \label{eq:detmvsr}
M_V^2 = {c_0 s_0^2 \over 4 F_\pi^2} - {c_2 \over 2 F_\pi^2}
= {3 \over N_c} \, 8 \pi^2 F_\pi^2 
\left( {1 \over 1 + {\bar\alpha_s \over \pi} } - 
       {2 \pi^2 N_c \over 9} \, 
       {\left\langle {\alpha_s \over \pi}G^2 \right\rangle \over (4 \pi F_\pi)^4 }
\right)   \,.
\end{equation}
Using (\ref{eq:mass1}) and (\ref{eq:gdet}) we finally find 
\begin{equation}
  \label{eq:detl3sr}
L_3 = - {F_\pi^2 \over 4 M_V^2} = -{1 \over 6} \, {N_c \over 16 \pi^2}
\left( {1 \over 1 + {\bar\alpha_s \over \pi} } - 
       {2 \pi^2 N_c \over 9} \, 
       {\left\langle {\alpha_s \over \pi}G^2 \right\rangle \over (4 \pi F_\pi)^4 }
\right)^{-1}
\end{equation}
and
\begin{equation}
  \label{eq:detgsr}
g^2 = - {1 \over 8 L_3} =
{3 \over N_c} \, 4 \pi^2 
\left( {1 \over 1 + {\bar\alpha_s \over \pi} } - 
       {2 \pi^2 N_c \over 9} \, 
       {\left\langle {\alpha_s \over \pi}G^2 \right\rangle \over (4 \pi F_\pi)^4 }
\right)   \,.
\end{equation}
Before comparing these results with the corresponding ones of the previous section
we should take a look at the third finite energy sum rule (\ref{eq:fesr3}):
Using (\ref{eq:coeffrho}) and (\ref{eq:pararho2})-(\ref{eq:detmvsr}) one obtains
\begin{equation}
  \label{eq:wrongfesr3}
-\left( {3 \over N_c} \right)^2 {128 \over 3} \pi^4 F_\pi^6
\left[ -3 
  \left( {1 \over 1 + {\bar\alpha_s \over \pi} } - 
         {2 \pi^2 N_c \over 9} \, 
         {\left\langle {\alpha_s \over \pi}G^2 \right\rangle \over (4 \pi F_\pi)^4 }
  \right)^2
  + {4 \over \left( 1 + {\bar\alpha_s \over \pi} \right)^2 }
\right]
\stackrel{?}{=}  -{112 \over 81} \pi \bar\alpha_s {3 \over N_c} \,
\langle \bar q q \rangle^2  \,.
\end{equation}
At first glance this looks like a clever way to express the quark condensate in
terms of the pion decay constant and gluonic corrections. However, we have already
expressed our doubts concerning the use of higher finite energy sum rules. The reason 
for this concern is their
higher sensitivity to the high-energy behavior which we have accounted for in an 
approximate way only. Indeed, plugging in reasonable numbers --- which will be
specified in the next section --- for the quantities appearing in (\ref{eq:wrongfesr3})
we get $-0.0026$ GeV$^6$ for the l.h.s.~and $-0.00033$ GeV$^6$ for the r.h.s., 
i.e.~almost a factor of ten difference. This indicates that at least the
third finite energy sum rule (\ref{eq:fesr3}) should not be trusted. An argument
why the other two sum rules are supposed to give more reliable results can only come 
from outside of the sum rule method, e.g.~by comparing the obtained results to the
quark model predictions of the previous section. This shows again the power of our
approach: We combine various methods to overcome the short-comings which appear
if only one method is used on its own.

The results (\ref{eq:detmvsr})-(\ref{eq:detgsr}) --- obtained from IAM + QCD sum 
rules --- can be compared to the 
results (\ref{eq:qmIAMmass}), (\ref{eq:quark3}) and (\ref{eq:qmIAMcoupl}) --- obtained
from IAM + quark model: Obviously the results agree up to
gluonic corrections which are neglected in the constituent quark model 
(cf.~the discussion
of that issue in \cite{espraf}). Due to the intrinsic uncertainties of the
QCD sum rule method discussed above we hesitate to take the results obtained in this
section as an improvement of the results (\ref{eq:qmIAM}) obtained in the last
section. We merely adopt the following point of view: Both approaches IAM + quark model
and IAM + QCD sum rules yield identical results up to gluonic corrections. We regard
this as the intrinsic uncertainty of the whole method. Hence the difference between 
the two approaches should be taken as a rough error estimate for the obtained results.
As we shall see in the next section this error turns out to be rather small.
We note in passing that the gluonic corrections in (\ref{eq:detl3sr}) have the same
sign as the one given in \cite{espraf}.

We would like to mention that the finite energy sum rules are not the only way
to use the dispersion relation (\ref{eq:opehadr}) in practice. The inventors of
the QCD sum rule method \cite{shif79} used Borel sum rules instead due to their lower
sensitivity to the high-energy parametrization (cf.~discussion above). We have checked
that the use of a Borel sum rule in the way described in 
\cite{Leupold:1998dg,Leupold:2001hj} leads to quantitatively similar results as compared
to the ones
presented here. We prefer to use the finite energy sum rules here since in this case 
the results can be presented in a closed analytic form.

\section{Comparing results with experiment and other models}
\label{sec:res}

We have obtained expressions for the mass of the $\rho$-meson, for its coupling to pions
and photons, for the low-energy coefficient $L_3$ and --- within the sum rule approach
--- for the continuum threshold $s_0$. Next we have to specify our input to allow for a 
comparison of the calculated
quantities with experiment. For the predictions from IAM + quark model we only need
the pion decay constant which we take as $F_\pi \approx 92\,$MeV \cite{pdg02}.
The calculations from IAM + sum rules require in addition the strong coupling constant,
the gluon condensate and the quark condensate. The latter we need only to show
that the third finite energy sum rule is invalid. The typical scale which appears in
the finite energy sum rules (\ref{eq:fesr}) is set by the continuum threshold $s_0$.
Consequently we evaluate the QCD running coupling constant at 
$\mu \approx \sqrt{s_0} \approx 4\pi F_\pi \approx 1 \,$GeV: 
In the following we use $\alpha_s(1\,{\rm GeV}) \approx 0.39$ \cite{GRV98}.
For the gluon condensate we use a canonical value \cite{shif79} of
$\left\langle {\alpha_s \over \pi} G^2 \right\rangle \approx (330 \,\mbox{MeV})^4$.
The quark condensate is determined from the Gell-Mann--Oakes--Renner relation \cite{GOR}
\begin{equation}
  \label{eq:gor}
m_q \langle \bar q q\rangle  = - {1 \over 2} F_\pi^2 M_\pi^2  \,.
\end{equation}
We use a value of $m_q(1\,{\rm GeV}) \approx 6\,$MeV for the average of up- and 
down-quark mass \cite{narison}. This yields 
$\langle \bar q q\rangle \approx -(240\,{\rm MeV})^3$. 

With these input values we find from quark model + IAM in the 
combined large-$N_c$ and chiral limit:
\begin{equation}
  \label{eq:numbersqm}
M_V \approx 820 \, {\rm MeV}\,, 
\qquad g \approx 6.3 \,,
\qquad L_3 \approx -3.2 \cdot 10^{-3}  \,.
\end{equation}
While from sum rules + IAM (in the same limit) we get:
\begin{equation}
  \label{eq:numberssr}
M_V \approx 760 \, {\rm MeV} \,,
\qquad g \approx 5.8 \,,
\qquad L_3 \approx -3.7 \cdot 10^{-3}  \,,
\end{equation}
and
\begin{equation}
  \label{eq:nums0}
s_0 \approx 1.2 \,{\rm GeV}^2  \,.
\end{equation}
We point out again that we do not regard the differences of the results 
(\ref{eq:numbersqm}) and (\ref{eq:numberssr}) as contradicting,
but merely as an estimate for the respective intrinsic uncertainties of both approaches.
Therefore we finally present the average values of corresponding results
and attribute as an error the difference of the average to the original results.
Of course this error can only be regarded as a rough estimate. We end up with
\begin{equation}
  \label{eq:numbersfin}
M_V \approx 790 \pm 30 \,{\rm MeV} \,,
\qquad g \approx 6.0 \pm 0.3 \,,
\qquad L_3 \approx (-3.4 \pm 0.3) \cdot 10^{-3}  \,.
\end{equation}
This has to be compared with the experimental results
\begin{equation}
  \label{eq:numbersexp}
M_V \approx 771 \,{\rm MeV} \,,
\qquad g \approx 6.05 \,,
\qquad L_3 \approx (-2.79 \pm 0.14) \cdot 10^{-3}  \,.
\end{equation}
where the $\rho$-meson mass is taken from \cite{pdg02}, the $\rho$-pion coupling
from \cite{klingl1} and the low-energy coefficient from the IAM fit to the scattering
data \cite{pelaez02}. We observe a very good agreement between (\ref{eq:numbersfin})
and (\ref{eq:numbersexp}). In addition, we find an error of 10\%
for $M_V^2$, $g^2$ and $L_3$, and consequently a 5\% error for $M_V$ and $g$.
The smallness of this error gives us further confidence that our approach is reasonable.
Finally we note that $s_0$ has to be between the 
masses (squared) of the $\rho$-meson and the next resonance in this channel, the $\rho'$.
The experimental average $(M_V^2+M_{\rho'}^2)/2 \approx 1.4\,$GeV$^2$ \cite{pdg02}
is reasonably close to the sum rule result (\ref{eq:nums0}).

The rest of this section is devoted to the comparison with related approaches.
As already pointed out in the introduction the new aspect of the work presented here
is the {\em combination} of various (already known) QCD related methods. 
Therefore it is hardly surprising
that parts of the presented techniques and results can be found also in other works. 
We shall comment here on the respective similarities and differences.

In \cite{klingl3,Marco:1999xz} a simplified version of the finite energy sum rules
(\ref{eq:fesr1}) and (\ref{eq:fesr2}) was used by neglecting all gluonic corrections,
i.e.~perturbative corrections as well as the gluon condensate. As an {\em input}
it was assumed that the continuum threshold $s_0$ is given by the typical scale of
chiral symmetry breaking, 
\begin{equation}
  \label{eq:thresweise}
s_0 = (4 \pi F_\pi)^2  \,.  
\end{equation}
Using the parametrization 
(\ref{eq:pararho}) for the spectral information 
and assuming the universality of the vector meson coupling the obtained results agree
with (\ref{eq:qmIAM}). Had we used the same simplification for the sum rules, 
i.e.~neglected gluonic corrections, the results of Sec.~\ref{sec:IAM} and \ref{sec:QCDSR}
indeed would have been in complete agreement with each other and with the ones of
\cite{klingl3,Marco:1999xz}. For the threshold we would have obtained 
(\ref{eq:thresweise}) instead of (\ref{eq:dets0sr}). 
Hence the results of \cite{klingl3,Marco:1999xz} are rather similar to ours. We note, 
however, that input and output are exchanged to some extent: In 
\cite{klingl3,Marco:1999xz} two sum rules are used (in a simplified version)
together with the input (\ref{eq:thresweise}) and $g=g_\gamma$ to obtain as an output
the determination of $M_V$ and $g$. The input choice are (plausible) assumptions.
In the approach presented here we used the IAM in the
large-$N_c$ limit to determine our input for the sum rules, namely (\ref{eq:pararho2}).
The output are expressions for $s_0$ and $M_V$ as given in (\ref{eq:dets0sr}) and
(\ref{eq:detmvsr}), respectively. The fact, that our output for $s_0$ is so plausible
that it is used as an input in \cite{klingl3,Marco:1999xz} demonstrates again the
intrinsic consistency of our approach. 

In \cite{Marco:2001dh} the approach of \cite{klingl3,Marco:1999xz} was extended 
to a simultaneous treatment of vector and axial-vector channel and by
including more resonances (cf.~(\ref{eq:ressum})). Two results of that analysis --- which
agreed reasonably well with recent data of $\tau$-decay --- should be mentioned here:
(i) the continuum thresholds for vector and axial-vector channel do not agree with each 
other, (ii) a meson which decays into $n$ pions couples (approximately) with a strength 
of $n \pi F_\pi^2$ to the (axial-)vector current. The latter rule is fulfilled by our
spectral information (\ref{eq:pararho2}) as the $\rho$-meson couples to two 
pions.\footnote{Note that the normalization conventions differ to some extent. We have 
translated the results to be in line with our conventions.}
From the same rule one obtains the result that the $a_1$-meson couples with a strength
of $3 \pi F_\pi^2$ to the axial-vector current.\footnote{The author thanks E.~Marco for
pointing that out to him.} We shall use these results to comment on another approach to 
determine the $\rho$-meson properties and the low-energy coefficients.

In \cite{golter} all {\em three} finite energy sum rules (\ref{eq:fesr}) are used
in the combined large-$N_c$ and chiral limit together with the corresponding ones for 
the axial-vector channel. Following \cite{weinb} it was {\em assumed} in addition
that the continuum thresholds for vector and axial-vector channel are the same ---
in contrast to the analysis of \cite{Marco:2001dh} --- and that $g=g_\gamma$ holds.
Using in total six sum rules together with the KSFR relation (\ref{eq:ksfr}) and the 
mentioned assumptions the authors of \cite{golter} consequently determined seven 
quantities, namely the $\rho$-meson mass (squared) (cf.~(\ref{eq:qmIAMmass})), 
\begin{equation}
  \label{eq:mrhogolter}
M_V^2 = {2 \sqrt{6} \over 5} \, 8 \pi^2 F_\pi^2 \approx 0.98 \cdot 8 \pi^2 F_\pi^2 \,,
\end{equation}
the strength of the $\rho$-meson contribution to the vector current (cf.~prefactor of
$\delta$-function in (\ref{eq:pararho2})),
\begin{equation}
  \label{eq:rhostrengthgolter}
2 \pi F_\pi^2  \,,
\end{equation}
the $a_1$-meson mass (squared), 
\begin{equation}
  \label{eq:ma1golter}
M_A^2 = 2 M_V^2  \,,
\end{equation}
the strength of the $a_1$-meson contribution to the axial-vector current,
\begin{equation}
  \label{eq:a1strengthgolter}
\pi F_\pi^2  \,,
\end{equation}
the continuum threshold common for both channels, 
\begin{equation}
  \label{eq:thresgolter}
s_0 = (4 \pi F_\pi)^2  \,,
\end{equation}
the gluon condensate,
\begin{equation}
  \label{eq:gluongolter}
\left\langle {\alpha_s \over \pi} G^2 \right\rangle = 
{384 \over 5}\, \pi^2 \, (5 - 2 \sqrt{6}) F_\pi^4  \,,
\end{equation}
and the quark condensate,
\begin{equation}
  \label{eq:quarkgolter}
\pi \bar\alpha_s {3 \over N_c} \, \langle \bar q q \rangle^2 =
{864 \over 25} \pi^4 F_\pi^6   \,.
\end{equation}
Note that the only scale which entered this analysis was provided by the strength of 
the pion contribution to the axial-vector current, i.e.~$\pi F_\pi^2$. Therefore all
dimensionful quantities where determined as multiples of appropriate powers of $F_\pi$.
In spite of the fact that this approach \cite{golter} is close in spirit to the one 
presented here and that some of the results are identical or close to our results we 
have the following concerns about the work presented in \cite{golter}: First of all,
we do not trust in the use of the third finite energy sum rule (\ref{eq:fesr3}) as
already pointed out above. It is actually the use of this sum rule together with
the corresponding one for the axial-vector channel which allows for the determination
of the condensates in terms of $F_\pi$. We doubt that such a simple relation as
(\ref{eq:gluongolter}) can be true, since two different physical phenomena come into play
here: The gluon condensate is caused by the scale anomaly while the pion decay constant
signals chiral symmetry breaking. Even if both phenomena have a common origin, as 
e.g.~stated in the instanton model \cite{diakpet}, it is hard to believe that the 
relation between the gluon condensate and the pion decay constant is as simple as 
(\ref{eq:gluongolter}). Concerning actual numbers the value for the quark condensate 
obtained from (\ref{eq:quarkgolter}) is rather high. In other words, the 
corresponding value for the quark mass obtained successively from (\ref{eq:gor}) would
be very low. Also the value for the $a_1$-strength (\ref{eq:a1strengthgolter}) is
questionable. It is a consequence of the assumption that the continuum threshold is the
same in both the vector and the axial-vector channel \cite{weinb}. According to the
recent analysis \cite{Marco:2001dh} the experimentally obtained strength is about
three times higher (cf.~discussion above). To summarize, the work presented in 
\cite{golter} is technically similar to the one presented here but there are important
aspects where the two approaches differ: (i) We regard the continuum with its
threshold parameter in (\ref{eq:pihadans}) as an approximation to the true
large-$N_c$ representation (\ref{eq:ressum}). The corresponding statement holds for
the axial-vector channel. There is no reason why the two thresholds should be the same.
(ii) We think that the parametrization (\ref{eq:pihadans}) is too crude to
fulfill the third sum rule (\ref{eq:fesr3}) which is more sensitive to the high-energy 
behavior as compared to the first two sum rules (\ref{eq:fesr1}) and (\ref{eq:fesr2}).
To utilize the third sum rule higher resonances should be taken into account.
(iii) We do not use the sum rules to relate the condensates to the pion decay
constant. In principle, this could be done also in our approach e.g.~by equating
(\ref{eq:qmIAMmass}) and (\ref{eq:detmvsr}). From our point of view, however, both
results are approximations. Therefore we use the gluon condensate to estimate the
error in the determination of the $\rho$-meson properties.

In \cite{eckgas,Ecker:1989yg} low-energy coefficients and mesonic resonances are
connected in a way different from the IAM presented here. There a chiral lagrangian with
Goldstone boson and resonance fields is proposed \cite{eckgas,Ecker:1989yg}:
\begin{equation}
  \label{eq:lagrressat}
{\cal L} = {\cal L}_1 + {\cal L}_{\rm kin} + {\cal L}_{\rm int}
\end{equation}
where ${\cal L}_1$ is given in (\ref{eq:chipt2}), ${\cal L}_{\rm kin}$ denotes the 
kinetic part for the resonance fields and ${\cal L}_{\rm int}$ the interaction part
between resonances and Goldstone bosons (and external sources).
By integrating out the resonances one arrives at the lagrangian of 
$\chi$PT (\ref{eq:chipt}) with predictions for the low-energy constants $L_i$ in terms
of resonance parameters, schematically 
\begin{equation}
  \label{eq:intres}
{\cal L}_{\rm kin} + {\cal L}_{\rm int} \to {\cal L}_2
\end{equation}
where ${\cal L}_2$ is given in (\ref{eq:chipt4}). We refer to this scheme as
``resonance saturation'' in the following. A reggeized version of that approach
was presented in \cite{Polyakov:1996vh}. A connection to a quark model is addressed in
\cite{Peris:1998nj}.
Note that the lagrangian ${\cal L}_1$ which mediates the
lowest-order interaction between the Goldstone bosons is already present in 
(\ref{eq:lagrressat}), i.e.~it is {\em not} generated by resonance exchange. 
This is in contrast to the IAM where the resonance properties are calculated from
a combination of $T_2$ and $T_4$. The former comes from a tree level calculation
using the lagrangian ${\cal L}_1$ while the latter obtains contributions from the
tree level of ${\cal L}_2$ and from the one-loop level of ${\cal L}_1$. Therefore
within the IAM both lagrangians ${\cal L}_1$ and ${\cal L}_2$ are connected to the 
resonance properties while in the approach presented in \cite{eckgas,Ecker:1989yg} this 
is only the case for ${\cal L}_2$. In spite of these differences it is interesting
to compare the results in the light of our analysis: Using scalar, pseudoscalar,
vector and axial-vector mesonic resonances it was shown in \cite{eckgas} that only 
(flavor octet) vector
mesons contribute to $L_1$ and $L_2$ while $L_3$ is additionally influenced by the 
exchange of scalar mesons. The results for $L_1$ and $L_2$ are
\begin{equation}
  \label{eq:l1l2ressat}
L_1 = {1 \over 2} L_2 = {G_V^2 \over 8 M_V^2}
\end{equation}
where $G_V$ denotes the coupling between vector mesons and Goldstone bosons. In our
language $G_V$ translates to 
\begin{equation}
  \label{eq:grhopipi}
G_V = {g F_\pi^2 \over M_V}   \,.
\end{equation}
Relation (\ref{eq:grhopipi}) can be obtained by calculating the decay width
$\Gamma(\rho \to \pi\pi)$ in both approaches (\ref{eq:lagrressat})
and (\ref{eq:standlagr}). Inserting (\ref{eq:grhopipi}) in (\ref{eq:l1l2ressat}) and
using the KSFR relation (\ref{eq:ksfr}) one gets
\begin{equation}
  \label{eq:l1l2ressat2}
L_1 = {1 \over 2} L_2 = {g^2 F_\pi^4 \over 8 M_V^4} = {1 \over 32 \, g_\gamma^2}  \,.
\end{equation}
Finally we can use the quark model expression for $L_1$ or $L_2$ given in 
(\ref{eq:quark1}) and (\ref{eq:quark2}), respectively. This yields again the relation
(\ref{eq:qmIAMcoupl}) for $g_\gamma$. Assuming in addition $g = g_\gamma$ finally 
allows to rederive all relations given in (\ref{eq:qmIAM}) \cite{Pich:1995bw}.
Hence the combination of IAM and quark model
on the one hand side and the combination of resonance saturation, quark model and 
universality of the vector meson coupling on the other hand side yield identical 
results for the $\rho$-meson properties.\footnote{This consistency has also been used 
in a preliminary version of the present work \cite{Leupold:2001vs} to derive 
(\ref{eq:qmIAM}).} In spite of this internal consistency it remains unclear how a
method which uses ${\cal L}_1$ and ${\cal L}_2$ (IAM) can be reconciled with a method 
which uses
only ${\cal L}_2$ (resonance saturation) to derive the same results. In that context it
is also interesting albeit unclear why it is $L_3$ which is connected to the $\rho$-meson
properties in the approach IAM + quark model, while it is $L_1 = L_2/2$ for the
approach resonance saturation + quark model. Clearly further work is needed to figure
out the relation between the IAM and the resonance saturation scheme.

We have used a very specific quark model in Sec.~\ref{sec:quarkmod} and we will
comment now on other related models. The lagrangian (\ref{eq:lagrconst})
can be regarded as the non-linear version of the standard Nambu--Jona-Lasinio (NJL) model
\cite{Vogl:1991qt,Hatsuda:1994pi,Klevansky:1992qe} (cf.~e.g.~discussion 
in \cite{Christov:1996vm}). While the quark model used here has only pions (in the SU(2)
sector), the standard NJL
model generates pions {\em and $\sigma$-mesons} as bound quark-antiquark states. 
In contrast to that it is easy to check that the IAM does not yield $\sigma$-mesons
in the large-$N_c$ approximation.\footnote{The full IAM as treated in \cite{pelaez02}
without further approximations reproduces the phase shifts of meson-meson scattering
rather well also in the scalar--iso-scalar channel. The fact that there is no pole
in the scattering amplitude in the large-$N_c$ approximation of the IAM 
indicates that the 
$\sigma$-meson is not a bound quark-antiquark state (see also the corresponding 
discussion in \cite{Oller:1998zr}).} Therefore we prefer to use the non-linear version
of the NJL model together with the IAM. 

Extended versions of the NJL model with additional quark-antiquark interactions in the 
vector channel have been used to generate the $\rho$-meson in
an RPA-type calculation \cite{Polleri:1997rw,He:1998gn}. This resembles the way how
the pion is generated from the standard NJL interaction 
term \cite{Vogl:1991qt,Hatsuda:1994pi,Klevansky:1992qe}. However, in contrast to the 
pion the $\rho$-meson mass is close to the production threshold for a constituent 
quark-antiquark pair. Therefore the spectral function of the $\rho$-meson and
correspondingly the scattering amplitude for pion-pion scattering is poisoned by
an unphysical inelasticity \cite{He:1998gn}. This is in contrast to the approach 
presented here where the quark model is only used in the low-energy regime much
below the unphysical production threshold. There the quarks are integrated out to
obtain the purely hadronic lagrangian of $\chi$PT with predictions for the chiral
low-energy coefficients. Hence the input for the IAM --- which extends the applicability
to the resonance region --- is purely hadronic. No unphysical quark-antiquark production
thresholds appear in the approach presented here. In addition we note that
the extended versions of the NJL model introduce an additional {\it a priori} unknown 
coupling constant which is chosen such that the physical value for
the $\rho$-meson mass is reproduced. In contrast, our approach
yields a parameter-free determination of the $\rho$-meson properties. No additional
interaction beyond the non-linear version of the standard NJL interaction is required.

In \cite{donhol} the calculation of the low-energy coefficient $L_{10}$ is addressed.
It is shown that $L_{10}$ can be obtained --- up to scale corrections --- as an integral 
over the difference of vector and axial-vector spectral functions 
\begin{equation}
  \label{eq:l10spec}
L_{10}(\mu) = -{1\over 4} \int \! {ds\over s} \, [\rho_V(s) - \rho_A(s)]
- {1 \over 144 \pi^2} [\log(m_\pi^2/\mu^2)+1]  \,.
\end{equation}
The authors compare two ways to obtain
these spectral functions: (i) from the Minkowski space version of the
constituent quark model (\ref{eq:lagrconst}) and (ii) from the experimental results
of $\tau$ decays. In the latter approach $\rho$- and $a_1$-peaks are 
clearly seen in the spectral functions. The approach (i)
of \cite{donhol} seems to be rather similar to the one presented here. However, there is
a subtle difference in the two approaches which has led the authors of \cite{donhol} to 
the rather negative conclusion that the constituent quark model is unreliable for the
calculation of the low-energy coefficients. In their approach they compared not only the
results for $L_{10}$ but already the spectral input, i.e.~the integrands of 
(\ref{eq:l10spec}) in both models (i) and (ii). Of course, in the quark model
the spectral functions are completely structureless showing no signs of $\rho$- and
$a_1$-peaks in contrast to the experimental data. Therefore the authors of \cite{donhol}
concluded that the quark model should not be used to calculate $L_{10}$. This conclusion
might be valid for the Minkowski space version of the constituent quark model
treated up to arbitrarily high energies. Our way of using the quark model is however
different. As already
discussed above, we prefer to use the quark model (a) in Euclidean space and 
(b) for low momenta only. Point (a) is more than
pure semantics. There are approaches which are only meaningful in the Euclidean region
and one specific example is the instanton model and all effective lagrangians derived
from it. Another example is the QCD sum rule approach discussed above. There, two
expressions for a correlation function are matched for {\em Euclidean} (space-like) 
momenta.
One expression is provided by an integral over a spectral function(!) while the other 
one is obtained from an operator product expansion (OPE). It is formally possible albeit 
completely misleading to calculate in Minkowski space the imaginary part of the OPE
expression and compare it to the spectral function. Both expressions look completely
different. The matching has to be performed in the Euclidean region. In this way it 
provides a powerful tool to learn something about the spectral function. 
Basically the same is true for the case at hand. (\ref{eq:l10spec}) can be used
for a model with hadronic degrees of freedom. However, according to our philosophy we 
would never use (\ref{eq:l10spec}) for the quark model. Of course, also point (b)
--- the restriction to low energies --- forbids the use of the quark model in the 
region of the hadronic resonances.
Therefore the negative conclusion drawn
in \cite{donhol} does not apply to the way in which we use the quark model.

\section{Summary and outlook}
\label{sec:sum}

To summarize we have calculated the mass of the $\rho$-meson and its coupling to
pions and photons in the chiral limit in leading order of the large-$N_c$ expansion.
To get these results we have combined three methods which can be derived from QCD
within well-defined approximations, namely the chiral constituent quark model 
\cite{diakpet,espraf},
chiral perturbation theory \cite{gasleut1,gasleut2} unitarized by the inverse amplitude 
method (IAM) \cite{Dobado:1997ps,pelaez02}, and QCD sum rules \cite{shif79}. 
Each of these methods
has specific short-comings which are avoided by combining the methods.
The obtained results
are in very good agreement with the corresponding experimental values. It is interesting
to note that the result for the $\rho$-meson mass contains only $F_\pi$ (up to gluonic
corrections). Therefore our approach suggests that the spontaneous breakdown of chiral
symmetry does not only determine the properties of the quasi-Goldstone modes but also
of the $\rho$-meson as the lightest non-Goldstone boson. While chiral symmetry breaking
quantitatively determines the $\rho$-meson properties, confinement enters in a
qualitative way: Using a quark model which does not have an explicit confinement
mechanism we have to work in the low-energy region which is sufficiently far away
from the unphysical constituent quark-antiquark production threshold. Confinement
demands to get rid off, i.e. integrate out the quarks to obtain a purely hadronic
effective lagrangian. Here the influence of quarks only appears 
in the coupling constants of the hadronic degrees of freedom. According to the 
confinement restriction only such a purely
hadronic lagrangian can be extended --- via the IAM --- to the region of interest, 
i.e.~the resonance region. 

As we have seen the
combined chiral and large-$N_c$ limit works rather well for the determination of hadronic
parameters. These parameters can be taken as an input for a purely hadronic model with 
pion and explicit resonance fields. Within that model one can take finite pion masses
and calculate e.g.~effects of the pionic cloud on the $\rho$-meson. Strictly speaking
these effects are suppressed by $1/N_c$, i.e.~of the same order as contributions
which were neglected for the derivation of the hadronic parameters. For the full
$\rho$-meson propagator this looks as following:
\begin{equation}
  \label{eq:fullrhoprop}
D_\rho^{-1}(s) = s- M_V^2 - \delta M_V^2(s) + \Pi(s)  \,,
\end{equation}
where $M_V^2$ is of order $N_c^0$ and determined in the present work while
$\delta M_V^2$ is of order $N_c^{-1}$ and neglected in the present work.
Both $M_V^2$ and $\delta M_V^2$ come from the quark-antiquark core of the physical
$\rho$-meson, whereas $\Pi$ is the self-energy of the $\rho$-meson caused by the
pion cloud. $\Pi$ is also of order $N_c^{-1}$ but its physical origin makes it different
from $\delta M_V^2$. 
In view of the promising results obtained in this work the
appearing picture is that $1/N_c$-corrections can be neglected at the quark level
but should be considered at the hadronic level, e.g.~to obtain a finite width for
the resonance. Obviously, at the present stage there
is no deeper justification for that point of view. 

In principle, the same reasoning holds for the chiral limit. There, however, it
is more straightforward how to relax that restriction. E.g.~one might treat up- and
down-quarks in the chiral limit but consider finite strange-quark masses. In such
an approach one can try to determine the properties of the $K^*$-resonances along the
lines presented here. Keeping finite quark masses, however, involves low-energy
constants beyond the ones presented in (\ref{eq:quark}). While the latter are
parameter-free, this fortunate feature does not hold for the former. Hence a stronger
model dependence will appear in the calculation of $K^*$ properties. Of course, one
can take the argument around and pretend to learn something more about the quark model.
In any case, this is beyond the scope of the present work.

Another possible extension of the work presented here concerns the in-medium 
modifications of the $\rho$-meson properties. At finite temperature the thermally
excited pions change the properties of the $\rho$-meson. In the framework of the
IAM this has been addressed in \cite{pelaeztemp}. However, in addition the
underlying quark structure might change at finite temperature. In other words, the
IAM can deal with a $\chi$PT lagrangian also at finite temperature. But whether the
low-energy coefficients of this lagrangian have an intrinsic temperature dependence
cannot be decided in a purely hadronic model. Again one can start with the chiral
constituent quark model and integrate out the quarks for the finite temperature case.
In this way one obtains a $\chi$PT lagrangian with temperature dependent coefficients.
This can serve as an input for the IAM to obtain the properties of $\rho$-mesons
at finite temperature. Work in this direction is in progress.

\acknowledgments The author thanks Ulrich Mosel for discussions and continuous support.
He also acknowledges stimulating discussions with Carsten Greiner, Eugenio Marco 
and Stefan Scherer.

\bibliography{literature}
\bibliographystyle{apsrev}

\end{document}